\documentstyle[prl,aps,psfig]{revtex}
\tighten
\twocolumn
\begin{document}
\draft
\preprint{\today}
\title{Critical quantum chaos and the one dimensional Harper model}

\author{S.N. Evangelou$^{(a)}$ and  J.-L. Pichard$^{(b)}$}
\address{(a) Department of Physics, University of Ioannina, Ioannina 45 110,
Greece}
\address{ (b) CEA, Service de Physique de l'\'Etat Condens\'e, 
Centre d'\'Etudes de Saclay, 91191 Gif-Sur-Yvette cedex, France}

\date{\today}
\maketitle

\begin{abstract}

We study the quasiperiodic Harper's model in order 
to give further support for a possible universality 
of the critical spectral statistics. 
At the mobility edge we numerically obtain a scale-invariant 
distribution of the bands $S$, which is closely 
described by a semi-Poisson $P(S)=4S \exp(-2S)$ curve. 
The $\exp (-2S)$ tail appears when the mobility edge 
is approached from the metal while $P(S)$ is 
asymptotically log-normal for the insulator. The multifractal 
critical density of states also leads to a sub-Poisson linear number 
variance  $\Sigma_{2}(E)\propto 0.041E$. 
 
\end{abstract}



 During the past few years the statistical description of the 
energy levels in  quantum systems has emerged as a new field
of study, bringing together the areas of mesoscopic physics
and quantum chaos \cite{guhr}. In disordered metals, the states 
are extended with correlated energy levels characterized 
by level-repulsion found in appropriate random matrix ensembles. 
This universal limit known as  Wigner statistics
also applies to quantum systems with chaotic classical 
dynamics \cite{bohigas}. In disordered insulators, the states 
are localized and the energy spectra obey  Poisson statistics, 
which is the generic limit for integrable systems. At the 
metal-insulator transition, another scale invariant limit 
applies for the spectral fluctuations \cite{shklovskii}, 
associated with the multifractal nature of the critical 
eigenstates, which are neither extended nor 
localized \cite{evangelou}. However, the analysis is 
complicated by boundary effects \cite{braun,kravtsov} which persist  
in the thermodynamic limit. But after taking specific boundary conditions,
or averaging over a combination of them, the critical statistics resembles 
that obtained in certain weakly chaotic quantum systems, 
such as pseudointegrable rational triangle billiards \cite{bogomolny}. 
Similar statistics also appears in certain many body spectra 
(e.g. two particles \cite{waintal} in a disordered chain with on-site 
interactions), where $P(S)$ is Wigner-like (linear) for small $S$ 
and Poisson-like (exponential) at large $S$, overall described 
by the semi-Poisson curve $P(S)=4S \exp(-2S)$.  This result is accompanied 
by a sub-Poisson linear number variance $\Sigma_{2}(E)\sim \chi E$ with
$\chi <1$.  Semi-Poisson $P(S)$ and sub-Poisson
$\Sigma_{2}(E)$ seem to be a common feature of different critical systems, 
raising the question of a possible universality. 
The semi-Poisson $P(S)$ can be exactly derived 
from a short range plasma model for the joint probability 
distribution of the energy levels, once the logarithmic pairwise 
interaction is truncated to nearest neighbours \cite{bogomolny}. 
Screened interactions were also found to be consistent with the non 
universal long range part of the spectral fluctuations in disordered 
metals \cite{jalabert,pichard} and characterize solvable models proposed 
by Gaudin \cite{gaudin} and Yukawa \cite{yukawa}. Our aim is to 
show that the signatures of ``critical chaos'' appear at the
mobility edge of the $1d$ Harper model, supporting the idea
of possible universality.
  
The quasiperiodic Harper's equation  \cite{aubry} 
has a mobility edge in one dimension with no Wigner statistics 
for the metal, due to the quasi-ballistic nature of the extended states.
The Poisson law is also never reached for the insulator,
due to the correlated localized states \cite {megann}. 
At the critical point the spectrum is multifractal \cite{tang} 
with gaps $G$ of all sizes distributed according to the inverse 
power-law $P(G)\sim G^{-(D+1)}$ and $D \approx 0.5$, the spectral
fractal dimension \cite{machida,geisel}. The distribution of gaps was 
interpreted as nearest-level spacing distribution characterized by 
strong level-clustering $P(G\to 0)\to \infty$, which is
very different from the level-repulsion $P(S\to 0)\to 0$ 
of the critical $P(S)$ \cite{shklovskii,braun}. 
This apparently contradicts the idea of universality of the critical 
fluctuations and led us to revisit the issue. 
In order to obtain a possibly universal behaviour, we should (i) 
unfold the fractal spectrum to get rid of the average variation of 
the density of states $\rho(E)$ by keeping the rescaled local 
mean level spacing $1/\rho(E)$ fixed,  (ii) identify the crucial role 
of boundary conditions \cite{braun,kravtsov}, 
and (iii) connect the critical level statistics 
to the observed multifractality \cite{chalker}. 
The difficulty of (i) lies on the wildly fluctuating 
multifractal spectrum and was circumvented in \cite{megann} 
where  $\rho(E)$ was obtained by mapping the system 
onto a ring threaded by a flux and averaging over all 
possible fluxes (or equivalently by choosing different boundary conditions 
in the chain ends). This could suggest to average over boundary conditions 
in order to achieve universality. Moreover, it leads us to argue that 
the distribution of bandwidths $S$ is more appropriate than 
the distribution of gapwidths $G$ 
for probing level statistics since we show that  
the normalized $P(S)$ at the mobility edge is close 
to the scale-invariant semi-Poisson curve 
observed in other critical systems \cite{bogomolny}.
On the basis of these findings  we 
suggest a certain universality of the critical fluctuations
and reveal their connection to multifractality.

We analyse  the Harper's equation
\begin{equation}
E \psi_n  =  \lambda \cos(2\pi\sigma n + \nu) \psi_{n}
+\psi_{n+1} +\psi_{n-1},
\end{equation}
where $\psi_n$ indicates the particle  wavefunction
of energy $E$ on site $n$, with $\nu=0$ \cite{aubry}. 
Eq. (1) for $\lambda=2$ also describes a particle in a two-dimensional 
lattice in a uniform magnetic field with $\sigma$ flux quanta 
per elementary plaquette. When $\sigma$ is an irrational number the 
period of the potential is incommensurate with the lattice period. 
We can consider generic irrationals which cannot be approximated 
``too well" by rationals taking $\sigma$  as the limit of successive 
rationals $M/N$, so that the potential becomes commensurate with the 
lattice with period $N$. Then the problem reduces to the diagonalization 
of $N\times N$ matrices with a spectrum  which consist of 
$N$ bands separated by $N-1$ gaps. This also
defines a scaling procedure where the incommensurate limit 
$N \to \infty$  becomes equivalent with the thermodynamic limit.
The states of Eq. (1) are extended when $\lambda <2$, with non-zero 
Lebesgue spectral measure $2|2-\lambda|$ independent of $N$. 
For $\lambda >2$  the states are localized
and the spectral measure falls off exponentially fast. The most 
interesting case is the critical point $\lambda=2$ where 
the spectrum and the wavefunctions are known to
have hierarchical multifractal structure 
\cite{tang,hiramoto,evangeloubis,siebesma}.
Thouless showed that the total band measure 
(sum of bandwidths) is proportional to $4.6649744644...N^{-1}$, 
which defines a universal number \cite{thouless}. The 
critical spectrum is multifractal which implies that 
the various bandwidths $S$ fall off with different power laws 
$N^{-1/\alpha}$, where $\alpha$ are singularity strengths having
density $f(\alpha)$. At the edges of the spectrum
the bands scale as $N^{-1/\alpha_{min}}$,
the  central band as $N^{-1/\alpha_{max}}$ 
and the mean band is $4.6649744644...N^{-1/\alpha_{0}}$, where 
$\alpha_{min}\approx 0.421,\alpha_{max}\approx 0.547$ and
$\alpha_{0}=0.5$ are the most prominent
spectral dimensions \cite{tang,hiramoto}.

We compute the bandwidths $S_{i}$ for long-$N$ chains
by finding the eigenvalues of the corresponding matrices
for periodic and anti-periodic boundary conditions using
the symmetry of the potential 
$\lambda \cos(2\pi\sigma n )$ in order to diagonalize 
the symmetric and antisymmetric tridiagonal matrices
suggested in \cite{thouless}. We find the
non-overlapping bandwidths $S_i$, $i=1,2,.., N$
and the gaps $G_i$,  $i=1,2,.., N-1$ for  the inverse golden mean 
$\sigma=(\sqrt(5)-1)/2$, which is the limit of the ratios 
of successive Fibonacci numbers $N$. 
At the mobility edge $\lambda=2$ the obtained
normalized bandwidth distribution $P(S)$ 
is displayed in Fig. 1 for various sizes $N$. It shows
linear level-repulsion and exponential tails,
overall being described by the semi-Poisson $P(S)=4S \exp(-S)$ 
curve, in sharp contrast with the  power-law 
$P(G)\sim G^{-3/2}$  obtained from  the distribution of 
gaps \cite{machida,geisel}. This suggests to examine whether 
the obtained $P(S)$, which gives the distribution 
of the energy shifts when the boundary conditions are twisted 
for a commensurate potential of periodicity $N$, 
can be also understood as a meaningful spacing 
distribution for the incommensurate limit.

\par
\vspace{.4in}
\centerline{\psfig{figure=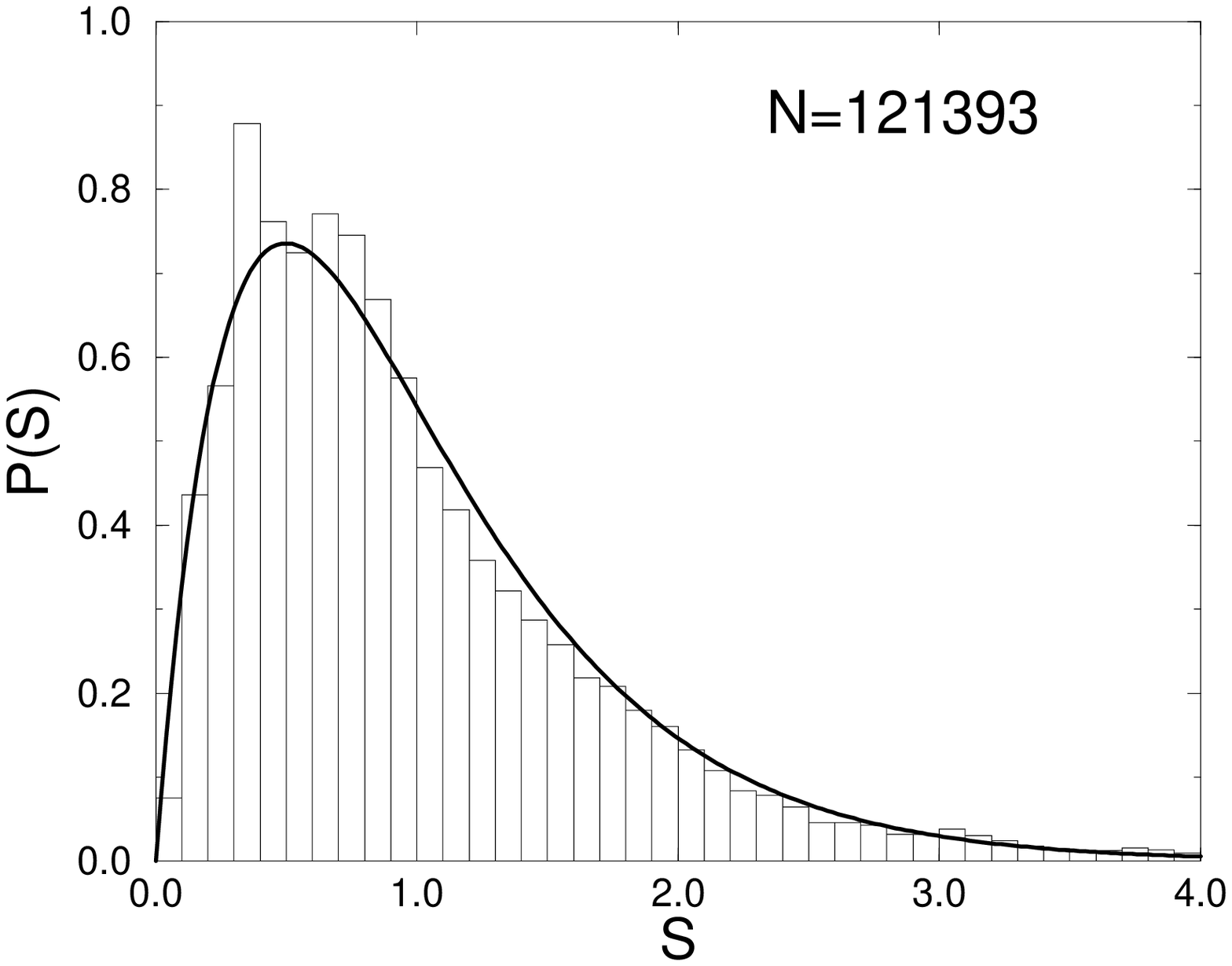,width=3.1in}}
\vspace{.2in}
\centerline{\psfig{figure=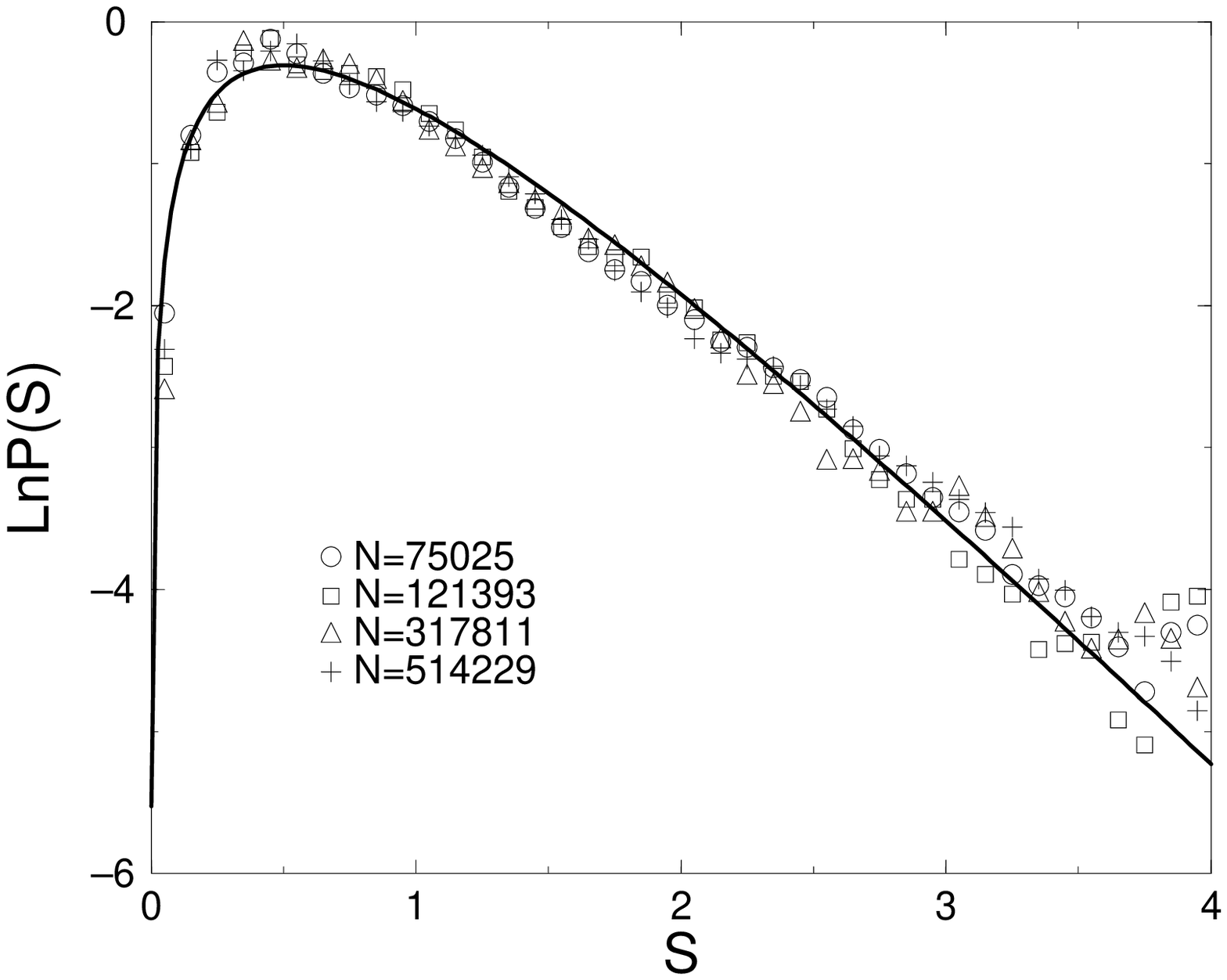,width=3.1in}}
{\footnotesize{{{\bf FIG. 1.} $\bf (a)$ The 
normalized nearest level spacing distribution
function at the mobility edge $\lambda=2$ 
for a characteristic size $N=121393$ (other sizes show
very similar pictures).
The solid curve is the semi-Poisson $P(S)=4S \exp(-2S)$.
$\bf (b)$ $P(S)$ in a semi-log plot for various sizes $N$ with the
semi-Poisson $\ln P(S)=\ln(4S) -2S$ result. 
The computed total band multiplied by the given
$N$ approaches the universal number of \cite{thouless}, 
within 4 to 8 decimals.}
}}
\par
The level fluctuations should be studied 
by considering the distribution of $(E_{i+1}-E_{i}) 
\rho(E_{i})$, where $\rho(E)$ is some averaged density 
of states at the energy $E$. 
In a disordered system this is usually done by averaging
over the ensemble at $E$ and in chaotic
systems one can use a semiclasical formula for 
the mean density \cite{bohigas}.
However, to unfold a multifractal
spectrum seems a hopeless task 
since the level spacing fluctuates violently 
making difficult to define
a mean  density. In other words one should
consider the statistics of $S/\langle G\rangle$  around $E$,
although the mean $\langle G\rangle$  cannot be defined. The solution 
was suggested in  ref. \cite{megann} which uses the unfolded levels 
from a mean $\rho(E)$ (or $\langle G \rangle$) obtained 
by repeating periodically the $N$-site chain or
changing the flux $\phi$, 
which describes the Bloch boundary condition
$\psi_{n+N}=\exp(i\phi) \psi_{n}$. 
A simpler way to do this is by noticing that the number of levels 
filling every band (which can be due to repeating
periodically the long chain of $N$ sites to infinity \cite{megann}),
is fixed ($\sim 1/N$) since to a given $\phi$ correspond $N$ levels,
one in every band. 
Moreover, each band has a smooth density of states
which can be made constant independently of its form,
by scaling the bandwidths $S_i$ with the fixed $N$. 
Thus, every bandwidth contributes a delta function to $P(S)$ 
centered at $S_i$, similarly to a delta function $P(S)$ 
at the mean spacing obtained for a smooth periodic band
and the level-spacing statistics can be
associated with the bandwidth distribution, 
essentially without unfolding. 
Viewed this way $P(S)$ concerns the distribution
of the mean spacings, each one of them arising
from the levels filling every band.
The shown scale-invariant $P(S)$ 
described by the semi-Poisson curve (Fig. 1) can also justify 
the results at the metal-insulator transition in three dimensional
disordered systems, where a similar $P(S)$ was obtained by averaging 
over boundary conditions \cite{braun,kravtsov}.

\par
A remarkable connection was recently made
between level-statistics and the multifractality
of critical wavefunctions \cite{chalker}. 
At the mobility edge the long range spectral fluctuations
are described by a linear number variance
$\Sigma_{2}(E)=\chi_{0}+\chi E$, with level compressibility 
$\chi = (1/2)(1-D^{\psi}_2 /d)$ related
to the correlation dimension $D^{\psi}_2$ of the 
critical wavefunctions and the space dimension $d$.
The quasiperiodic  model has no unique  wavefunction 
exponent $D^{\psi}_2$  to describe the compressibility, 
since $D^{\psi}_2$  varies
with the critical energies. In the Harper's model
$\Sigma_{2}(E)$ is directly influenced by the spectral fluctuations 
summarized in the presence of the multifractal
dimensions $D_{q}, q\in (-\infty, +\infty)$ 
and their Legendre transform $\alpha-$ $f(\alpha)$
\cite{tang,hiramoto}.  They are locally  defined 
around the spectral energy $E_{in}$, via
\begin{equation}
{\cal N} (E_{in}+E) - {\cal N} (E_{in}) \sim E^{\alpha},
\end{equation}
where ${\cal N} (E)$ is the number of levels 
from the  lowest energy up to $E$. 

\par
\par
\vspace{.4in}
\centerline{\psfig{figure=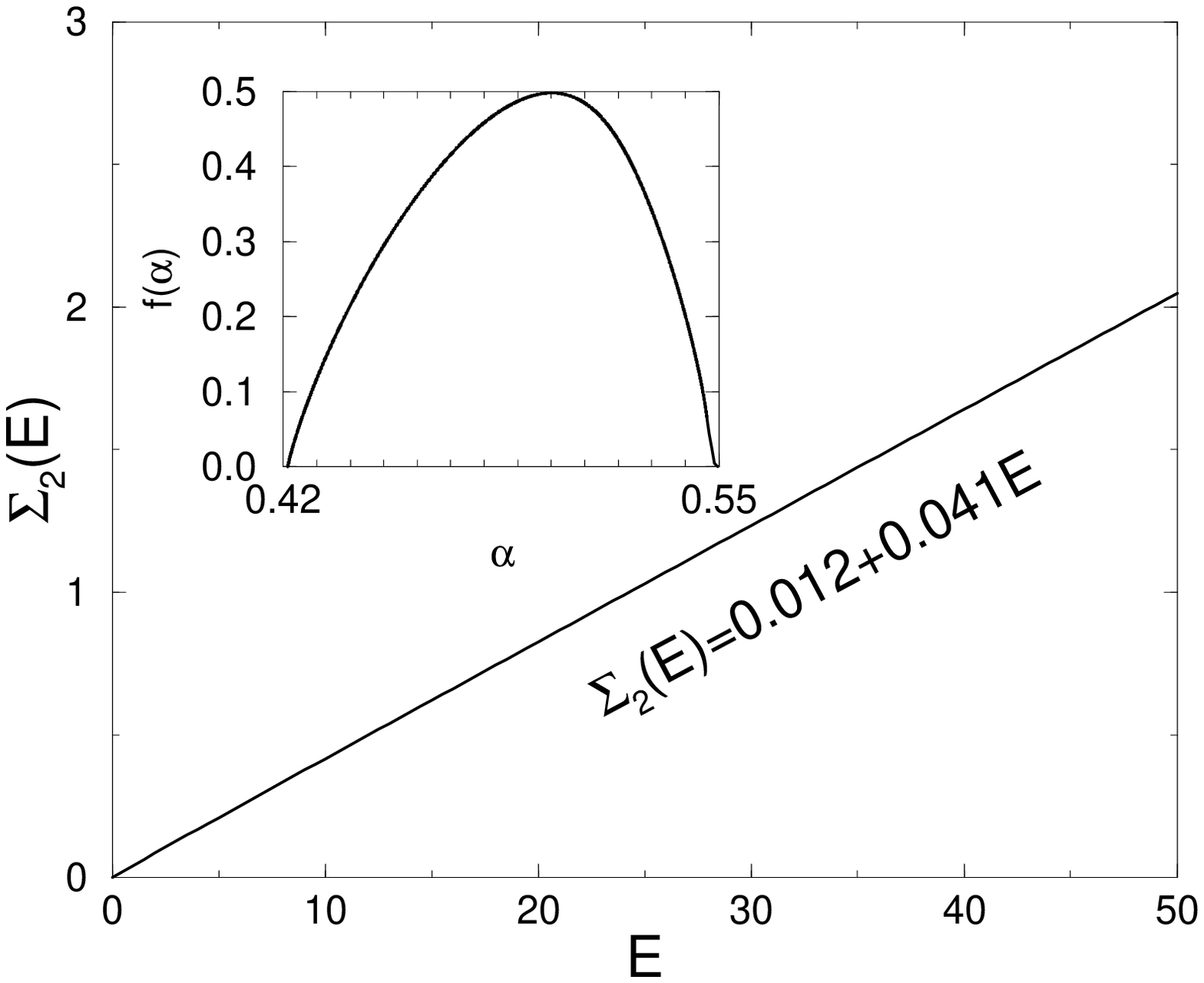,width=3.1in}}
{\footnotesize{{{\bf FIG. 2.} The 
number variance at the mobility edge $\lambda=2$.
Inset: the $\alpha$-$f(\alpha)$ spectral
dimensions.}
}}
\par
In the formalism of multifractals a spectral scaling 
function $\tau (q)$ is usually evaluated by 
box counting in energy, where an exact partioning of boxes is
suggested by the bandwidths $S_i$ since no levels fall in the
gaps. The  scaling dimensions are obtained by scaling
the moments of the computed distribution of $S_{i}$ \cite{tang} via
\begin{equation}
\sum_{i=1}^{N} S_{i}^{-\tau(q)} \sim N^{q}.
\end{equation}
For up to $N=313897$ it gives the spectral exponents 
$D_{q}=\tau(q)/(q-1)$ with $D_{-1}=0.5$, $D_0=D\approx 0.498$, 
$D_1\approx 0.496$, $D_2\approx 0.493$, etc.  
Their  Legendre transform  gives
the singularities $\alpha ={\frac {d\tau(q)}{dq}}$ 
and their density $f(\alpha)=q\alpha-\tau(q)$ with
the extreme dimensions $D_{+\infty}=\alpha_{min}$,
$D_{-\infty}=\alpha_{max}$.
The position of the maximum lies at 
$\alpha_{0}=0.5$ with $f(\alpha_{0})=D$
(see inset of Fig. 2). 

%
%
\par
\vspace{.4in}
\centerline{\psfig{figure=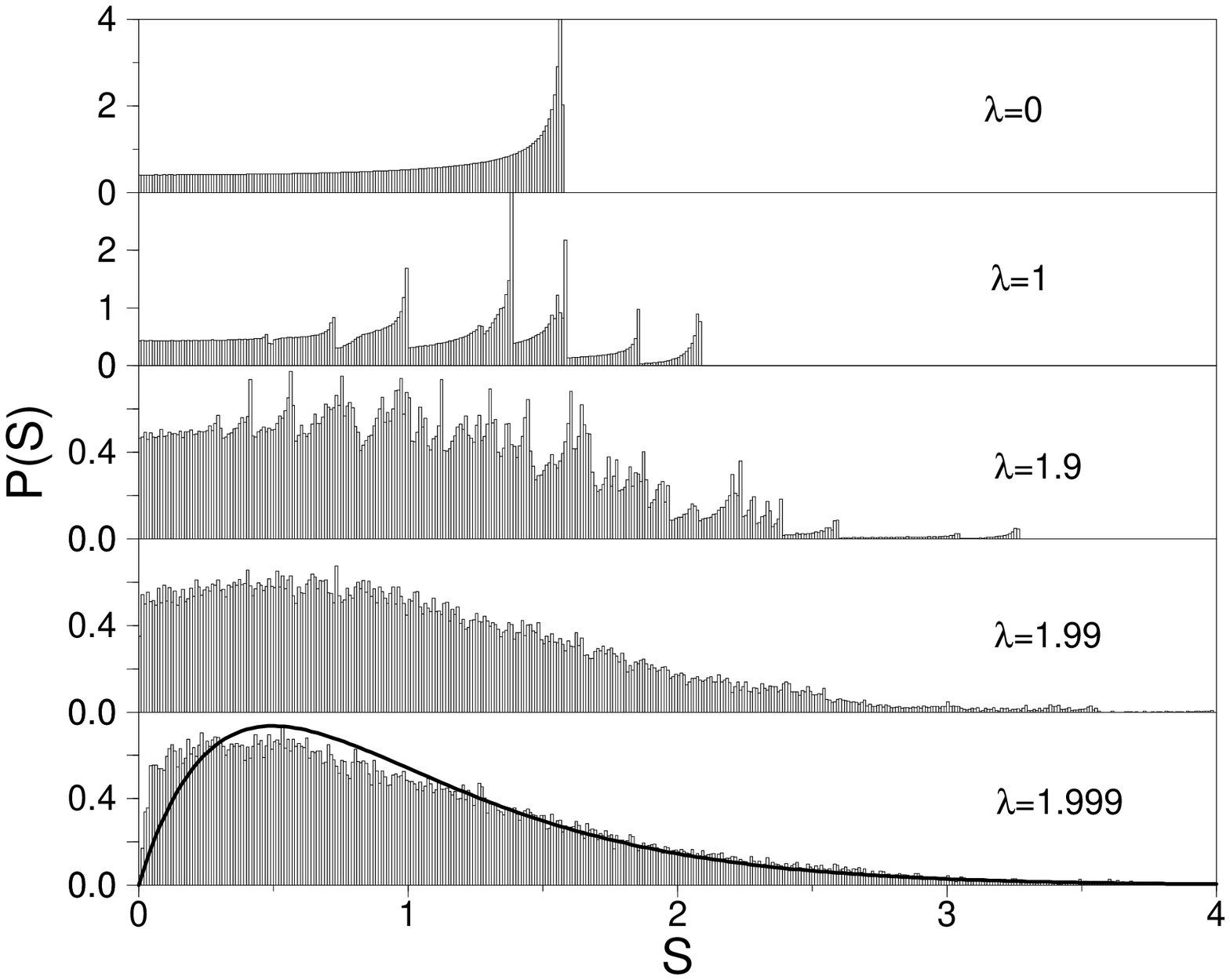,width=3.1in}}
{\footnotesize{{{\bf FIG. 3.} The 
nearest level spacing distribution
function in the metallic phase  $\lambda<2$ for $N=121393$.
The continuous line is the semi-Poisson limit approached when
$\lambda=2$.}
}}

\par
The level number variance 
\begin{eqnarray}
\Sigma_{2}(E) &  =  & {\int dE_{in}
 ({\cal N}(E_{in}+E) - {\cal N} (E_{in}))^{2}} \nonumber \\
 &  & - \left ({\int dE_{in}
 ({\cal N}(E_{in}+E) - {\cal N} (E_{in})})\right )^{2},
\end{eqnarray} 
is defined by taking energy boxes of size $E$ 
over all possible $E_{in}$ in the spectrum.  
In the box multifractal formalism the 
knowledge of $\alpha-f(\alpha)$ makes Eq. (4) 
strictly equivalent to
\begin{equation}
\Sigma_{2}(E)  =   {\int_{\alpha_{min}}^{\alpha_{max}} 
d\alpha f(\alpha) E^{2\alpha}}-\left 
({\int_{\alpha_{min}}^{\alpha_{max}}d\alpha f(\alpha)
E^{\alpha}}\right)^{2},
\end{equation}
which is approximately linear $\Sigma_2 \approx 0.012+0.041E$ as
shown in Fig. 2, leading to level-compressibility $\chi\approx 0.041$.
The obtained $\chi$ should be contrasted with the higher values 
$\chi\approx 0.10, 0.125$ for $d=2$ and $\chi\approx 0.32$ 
for $d=3$ critical disordered systems \cite{chalker}. 
In the one-dimensional mobility edge
$\chi$ is smaller, inversely proportional to  $d$
as expected from the formula for $\chi$
via the wavefunction exponent $D_2^{\psi}$ \cite{chalker}. 
The computed value of $\chi$ applied to this formula
gives $D_2^{\psi}\approx 0.918$, which can be compared with 
$D_2^{\psi}\approx 0.82$ obtained for states near the
band center \cite{siebesma}.  Since all the levels in the band 
correspond to critical states with variable $D_2^{\psi}$
we have also computed directly
the average exponent for all critical states which gave
the much lower value $D_2^{\psi}\approx 0.537$. 

\par
The level statistics of the Harper's equation
for $\lambda <2$ or  $\lambda >2$ can be simply understood
from a duality argument \cite{aubry}. In the metallic regime
we follow $P(S)$ for increasing
$\lambda$ in Fig. 3. The Poisson tail $\exp(-2S)$ is seen to emerge 
when approaching the critical point $\lambda=2$.
The spectral  dimensions in
the metallic limit are simply $D_{q}=1$, for $q<2$, and
$D_{q}=1/(2-2/q)$, for $q\ge 2$.
For example, for $\lambda=0$ we trivially 
obtain $S_{j}= (2\pi/N) |sin(\pi j M/N)|, 
j=1,2,..N$, which leads to $P(S)= {\frac {2} 
{\pi \sqrt{(\pi/2)^{2}-S^{2}}}}$ shown in Fig. 3, with
only one singularity $\alpha=0.5$, $f(\alpha=0.5)=0$.
More and  more singularities develop as we approach the
critical point $\lambda=2$ where $\alpha$ lies in
$(\alpha_{min}, \alpha_{max})$ with finite density $f(\alpha)$. 
The obtained critical semi-Poisson $P(S)$ gives another hint
towards the fascinating problem of the
analytical computation of 
the spectral multifractal dimensions, possibly from finite-$N$ 
corrections to strings in the recent formulation
via Bethe-anzatz equations \cite{abanov}.
In the localized regime $\lambda >2$ the bands 
become exponentially small and $P(S)$ 
asymptotically log-normal since the localized 
levels  are spatially correlated.
In the extended phase almost all  $\alpha=1$ and  
in the localized phase $\alpha =0$, which lead to
trivial $\Sigma_2(E)=0$  in the two limits.

The  main result of our study, which is  
able to treat numerically very long chains,
is the appearance of a semi-Poisson $P(S)=4S \exp(-2S)$
distribution  at the mobility edge in one dimension.
This associates the mobility edge with level-repulsion
for small spacings, rather than level-clustering, 
in agreement with $3d$ critical disordered systems. 
At the mobility edge
both the multifractal spectrum and the nearest-level statistics  
are seen to arise from the semi-Poisson distributed bandwidths $S_i$,
whose moment scaling describes  a fractal spectrum 
with infinite many singularities $\alpha -f(\alpha)$.  
A second result directly obtained from the multifractal spectrum
is a sub-Poisson linear number variance 
$\Sigma_{2}(E) \sim \chi E$, 
in addition to the approximate validity of the proposed formula 
for $\chi$ in terms of the multifractal wavefunctions.  
In conclusion, the obtained universal critical statistics
for the Harper's model
verifies numerically the ``critical chaos" scenario 
summarized in the semi-Poisson $P(S)$ and 
the sub-Poisson linear number variance $\Sigma_{2}(E)$.
``Critical chaos" is seen to be intimately connected 
to spectral and wavefunction multifractality.
Our study can also shed light on the average over boundary conditions
in the critical statistics of disordered systems.

This work was supported by a TMR of the EU
and a $\Pi$ $E$ $N$ $E$ $\Delta$ grant from
the Greek Secretariat of Science and Technology. 
S.N.E. is grateful for the hospitality at Saclay where this work 
was initiated.

\end{document}